\documentclass[aps,prb,twocolumn,groupedaddress,eqsecnum,%
showpacs,amsmath,%
floatfix%
]{revtex4}

\usepackage{graphicx}
\usepackage{dcolumn}
\usepackage{bm}

\begin{document}

\title{Phonon-assisted relaxation kinetics of statistically-degenerate
excitons \\ in high-quality quantum wells}
\author{A.V. Soroko}
\email[e-mail: ]{Alexander.Soroko@astro.cf.ac.uk}
\author{A.L. Ivanov}
\affiliation{ Department of Physics and Astronomy, Cardiff University,
Cardiff CF24 3YB, Wales, United Kingdom }
%
%
\begin{abstract}
    Acoustic-phonon-assisted thermalization  kinetics of  excitons
in quantum wells  (QWs) is developed  for small concentrations  of
particles,  $\rho_{\rm  2D}  \lesssim  10^9$ cm$^{-1}$, when
particle-particle interaction can be neglected while Bose-Einstein
statistics already strongly influences the relaxation processes at
low  temperatures.    In  this  case thermalization of QW excitons
occurs  through   nonequilibrium  states  and  is given by the
following scenario.  During the first transient stage, which lasts
a few  characteristic scattering  times, the  correlations with an
initial  distribution  of  QW  excitons  disappear.    The  next,
adiabatic   stage   of    thermalization   usually   takes    many
characteristic scattering times, depends only upon  two control
parameters,  the  lattice  temperature  $T_b$  and  the degeneracy
temperature $T_0 \propto \rho_{\rm 2D}$, and is characterized by a
quasi-equilibrated distribution  of high-energy  QW excitons  with
effective temperature $T(t)$.  We show that the thermalization law
of high-energy  particles is  given by  $\delta T(t)  = T(t) - T_b
\propto e^{-\lambda_0 t}/t$, where $\lambda_0$ is a marginal value
of the continuous eigenvalue spectrum of the linearized  kinetics.
By   analyzing   the   linearized   phonon-assisted   kinetics  of
statistically-degenerate  QW  excitons,  we  study  the dependence
$\lambda_0 =\lambda_0(T_b,T_0)$.  Our numerical estimates refer to
high quality  GaAs and  ZnSe QWs.   Finally,  we propose a special
design of  GaAs-based microcavities,  which considerably  weakens
the bottleneck effect in  relaxation of excitons (polaritons)  and
allows us to optimize the acoustic-phonon-assisted  thermalization
processes.
\end{abstract}
%
\pacs{78.66-w, 72.10.Di, 63.20.Kr}

\maketitle

\section{INTRODUCTION}
\label{Introduction}

    The formation, resonant or phonon-assisted, of QW excitons and
their following  relaxation towards  a final  (quasi-) equilibrium
thermodynamic  state  at  low  lattice  temperature  $T_b$ are the
subject  of  numerous  experimental  \cite{Shah96} and theoretical
\cite{Takagahara85,Vinattieri94,Selbmann96,Gurioli98,Gulia97,Oh00}
studies.    Recently,   relaxation  thermodynamics  has   been
formulated and  developed in  order to  analyze how  Bose-Einstein
statistics   of   high-density   QW   excitons   influences    the
phonon-assisted thermalization processes.  \cite{Ivanov99PRB}  The
above thermodynamics  assumes a  strong, dominant  exciton-exciton
scattering and, therefore, relaxation through  quasi-equilibrium
thermodynamic   states.      The   relaxation  thermodynamics  has
successfully been applied  to model the  $\rho_{\rm 2D}$-dependent
thermalization  and  photoluminescence  kinetics  observed  in 
early experiments \cite{Damen90,Kuhl91} with high-density excitons
($5   \times   10^9$ cm$^{-2}$ $   \leq   \rho_{\rm   2D}    \leq
10^{11}$ cm$^{-2}$) in GaAs QWs.  \cite{Ivanov99PRB}

    The long-lived indirect  excitons in high-quality  GaAs/AlGaAs
coupled  QWs  provide  a  unique  opportunity for studying quantum
degeneracy in a system of  two-dimensional bosons.  In this  case,
the  long  radiative  lifetimes  of  indirect  excitons allows the
system to cool down to  temperatures where the dilute exciton  gas
becomes    statistically     degenerate. 
\cite{Fukuzawa90,Butov94,Butov98,Butov99,Butov01,Larionov2001,Krivolapchuk2001}
The  quality  of  present  day  GaAs/AlGaAs  coupled  QWs has been
considerably improved in comparison with those used in the pioneering
experiments  
\cite{Fukuzawa90,Charbonneau88,Kash90,Golub90,Golub92,Deveaud90} 
one decade ago.  Furthermore, the very  recent
magneto-optical experiments  \cite{Parlangeli00} clearly  indicate
that  the  in-plane  momentum  $\hbar  {\bf  k}_{\|}$  of indirect
excitons  is  a   well-defined  quantum  number   in  high-quality
GaAs/AlGaAs coupled QWs.

    Thermalization  of  hot  photoexcited  excitons  down  to  the
temperature of the  cold lattice occurs  mainly via scattering  by
thermal bulk longitudinal acoustic  (LA) phonons and is  much more
efficient for quasi-two-dimensional (quasi-2D) systems as compared
to  bulk  semiconductors.    This  follows  from the relaxation of
momentum  conservation  in   the  $z$-direction  (the   QW  growth
direction)  for  quasi-2D  systems:    the ground-state mode ${\bf
k_{\|}}={\bf 0}$,  i.e., the  energy state  $E=0$, couples  to the
continuum energy states  $E \geq E_0$,  rather than to  the single
energy state $E =  E_0 = 2 M_x  v_s^2$ ($v_s$ is the  longitudinal
sound velocity  and $M_x$  is the  in-plane translational  mass of
excitons) as occurs in bulk materials.  As a result, the LA-phonon
assisted kinetics  of QW  excitons becomes  dominant at $\rho_{\rm
2D}  \lesssim  1  -  3 \times10^9$  cm$^{-2}$:     in  this   case
exciton-exciton scattering  can be  neglected while  Bose-Einstein
(BE) statistics already strongly influences the relaxation process
at low temperatures.  \cite{Ivanov99PRB} Crossover from  classical
to quantum statistics occurs near the degeneracy temperature  $k_B
T_0 = 2 \pi \hbar^2 \rho_{\rm 2D}/(g M_x)$, where $g$ is the  spin
degeneracy factor.  For $\rho_{\rm 2D} = 3 \times 10^9$  cm$^{-2}$
the  degeneracy  temperature  of  indirect excitons in GaAs/AlGaAs
coupled QWs  is $T_0  = 0.79$  K. This  estimate refers  to $g=1$,
which can be  achieved in the  $b$-type GaAs/AlGaAs coupled  QWs by
applying  a  static   magnetic  field  ${\bf   H}  \|  {\bf   z}$.
\cite{Butov98} Note that the very recent experiments \cite{Butov01}
deal with GaAs/AlGaAs coupled QWs at extremely low cryostat
temperature $T_b=0.05$ K. 

    In this paper we study analytically and model numerically  the
acoustic-phonon-assisted       relaxation       kinetics        of
statistically-degenerate QW excitons at low densities.  The recent
experiments  \cite{Umlauff98,Bacher98}  allow  us  to visualize, by
means  of   LO-phonon-assisted  emission,   the  LA-phonon-assisted
kinetics of quasi-2D excitons in  ZnSe QWs and, in particular,  to
prove that for $\rho_{\rm  2D} \lesssim 10^9$ cm$^{-2}$  the above
kinetics  indeed  occurs  through  non-equilibrium distributions of QW
excitons.                    In                     experiments
\cite{Butov94,Butov98,Butov99,Butov01}   the   concentration    of
BE-degenerate indirect excitons in GaAs/AlGaAs coupled QWs usually
varies from $\rho_{\rm 2D} \gtrsim 10^{10}$ cm$^{-2}$ at the  very
end  of  an  optical  excitation  pulse to $\rho_{\rm 2D} \lesssim
10^8$ cm$^{-2}$ at large delay  times $t \gtrsim 50$ ns  after the
optical excitation.  Thus, thermalization of the indirect excitons
at large  delay times  cannot be  described within  the relaxation
thermodynamics  \cite{Ivanov99PRB}  and   does  need  a   separate
theoretical analysis.

    The classical Boltzmann kinetic equation has been  generalized
in order to include  quantum statistics by Uehling  and Uhlenbeck.
\cite{Uehling32}  The  relevant   quantum  kinetic  equation   for
spatially-homogeneous  dilute  system  of statistically-degenerate
quasi-2D excitons coupled to bulk LA-phonons is
\begin{widetext}
\begin{eqnarray}
\frac{\partial}{\partial t} N_{\bf k_{\|}} = &-&\frac{2 \pi}{\hbar}
\sum_{\bf q} |M(q,q_z)|^2 \Big\{ \Big[ N_{\bf k_{\|}}
(1+n^{\rm ph}_{\bf q})
(1+N_{{\bf k_{\|}}-{\bf q}_{\|}})
- (1+N_{\bf k_{\|}}) n^{\rm ph}_{\bf q}
N_{{\bf k_{\|}}-{\bf q}_{\|}} \Big]
\delta(E_{\bf k_{\|}} - E_{{\bf k_{\|}}-{\bf q}_{\|}} -
\hbar q v_s)
\nonumber\\
&+& \Big[ N_{\bf k_{\|}} n^{\rm ph}_{\bf q}
(1+N_{{\bf k_{\|}}+{\bf q}_{\|}})
 -(1+N_{\bf k_{\|}}) (1+n^{\rm ph}_{\bf q})
N_{{\bf k_{\|}}+{\bf q}_{\|}} \Big]
\delta(E_{\bf k_{\|}} - E_{{\bf k_{\|}}+{\bf q}_{\|}}
+ \hbar q v_s) \Big\} ,
\label{1.1}
\end{eqnarray}
\end{widetext}
    where  $N_{\bf  k_{\|}}$  and  $n^{\rm  ph}_{\bf  q}$  are the
occupation numbers  of exciton  in-plane mode  ${\bf k_{\|}}$  and
phonon  bulk  mode  ${\bf  q}=\{{\bf q_{\|}},q_z\}$, respectively,
and ${\bf q}_{\|}$ is the in-plane  projection of ${\bf q}$.
The terms  in the  first and  second square
brackets  on  the  right-hand-side  (r.h.s.)  of Eq.~(\ref{1.1})
describe the Stokes and anti-Stokes LA-phonon-assisted  scattering
processes, respectively.  The acoustical phonons are assumed to be
in   a   thermal   equilibrium at the bath   temperature    $T_b$.
\cite{Ivanov99PRB}  The  matrix  element  is  given by $M(q,q_z) =
[(D_x^2 \hbar q)  / (2 \rho  v_s V)]^{1/2} F_z(q_z  L_z/2)$, where
$\rho$  is  the  crystal  mass  density,  $D_x$ is the deformation
potential  of  exciton  -  LA-phonon  interaction, 
$L_z$ is the thickness of a QW, and $V$ is the
volume.   The form-factor  $F_z(\chi) =  [\sin(\chi)/ \chi]  [e^{i
\chi} /  (1 -  \chi^2/\pi^2)]$ refers  to an  infinite rectangular
confinement potential.   \cite{Bockelmann94}  The latter  function
describes the relaxation of  the momentum conservation law  in the
$z$-direction   and   characterizes   a   spectral  band  of  bulk
LA-phonons, which  effectively interact  with QW  excitons.   Note
that  Eq.~(\ref{1.1})  is  valid  only  for  the kinetic stage of
thermalization, i.e., before a low-temperature collective state of
excitons \cite{Lozovik75,Littlewood95,Littlewood96,Naveh96} builds
up.

\begin{figure}
\vspace{2cm}
\includegraphics{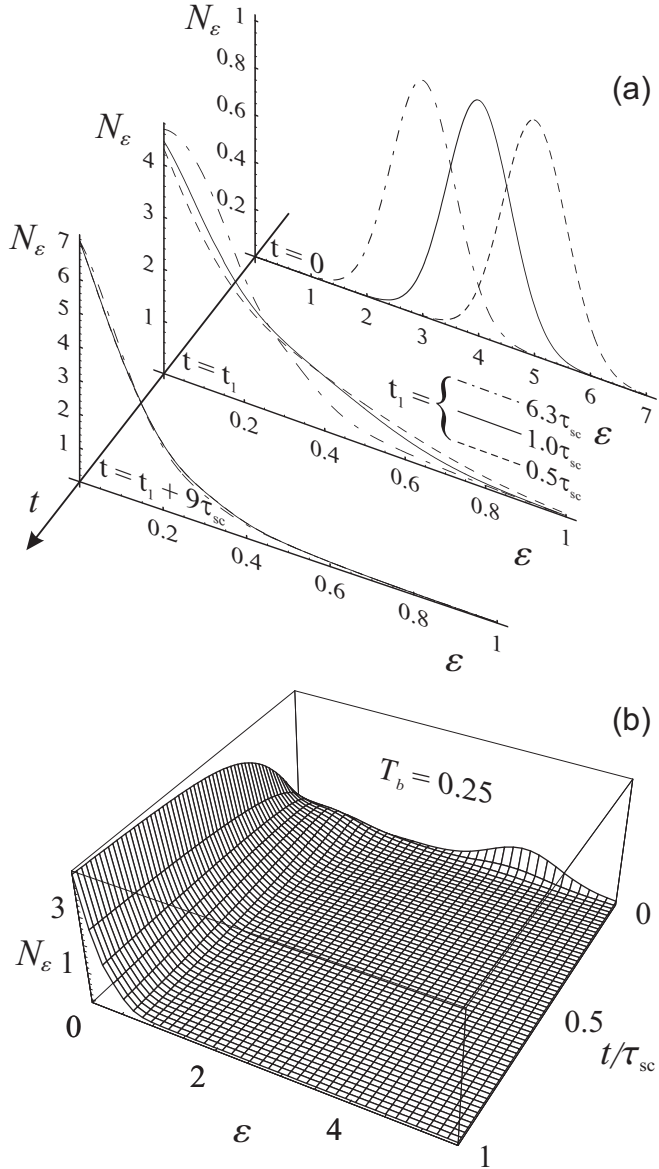}
\caption{
    (a)  Transient  relaxation  towards  the  adiabatic  stage  of
evolution calculated  for various  initial Gaussian  distributions
$N_{\varepsilon}(t=0)  \propto  e^{-  15.625  (\varepsilon - {\bar
\varepsilon})}$,   where 
$\varepsilon= \hbar^2 {\textbf k^2_{\|}}/(2 M_x E_0)$, and
parameter   $\bar   \varepsilon   =  3$
(dash-dotted lines),  $4$ (solid  lines), and  $5$ (dashed lines).
The  gross  dependence  of  evolution  upon  initial conditions is
absorbed  by  time  $t_1  \simeq  t_c$.    (b)  The first stage of
LA-phonon-assisted  relaxation  ($t  \lesssim  \tau_{\rm  sc}$)  for  a
particular   distribution   of   indirect   excitons   with  $\bar
\varepsilon = 4$.  In  both plots $T_b=0.25$ and $T_0=1.14$.   The
dimensionless    values,    time    in  $\tau_{\rm    sc}$   and
energy/temperature  in  $E_0$,  can  easily  be  rescaled  to  the
dimensional units  by using  $E_0=33$ $\mu$eV  and $\tau_{\rm sc}=
41$ ns for GaAs/AlGaAs coupled QWs, 
and $E_0=162$ $\mu$eV and $\tau_{\rm sc}= 2.7$ ns relevant to ZnSe QWs.
}
\label{fig1}
\end{figure}

    The main aim of our work is to study the fundamental  features
of  the  acoustic  phonon-assisted  thermalization  kinetics of QW
excitons    from    initial    strongly   nonequilibrium   $N_{\bf
k_{\|}}(t=0)$  towards  the  final  equilibrium  distribution with
well-developed Bose-Einstein statistics, when $N^0_{\bf  k_{\|}=0}
\gtrsim 1$.  Our  numerical simulations of the  LA-phonon-assisted
kinetics  clearly  demonstrate  that  after the first transient, which
lasts a few characteristic scattering times, 
a slow \textit{adiabatic stage} 
of thermalization builds  up (see Fig.~\ref{fig1}).   This
stage  is  characterized  by  a  quasi-equilibrium distribution of
high-energy    QW    excitons    with    effective     temperature
$T(t)=T_b+\delta  T(t)$   and  is   independent  of   the  initial
distribution  at   $t=0$.     The  adiabatic   stage  lasts   many
characteristic  scattering  times  and  arises  due to the need to
populate the low-energy in-plane modes with $N_{\bf k_{\|}\simeq0}
\gtrsim  1$  in  the  presence  of  effective  suppression  of  the
stimulated scattering processes (e.g., an intense incoming, Stokes
flux of excitons  into the ground-state  mode ${\bf k_{\|}}=0$  is
nearly compensated by the outgoing, anti-Stokes scattering out  of
the state ${\bf k_{\|}}$=0).   In order to describe the  adiabatic
stage  of  relaxation,  we find a \textit{generic solution} of the
quantum kinetic Eq.~(\ref{1.1}).   While the generic solution  we
calculate     is     different     from     that     derived    in
Ref.~\onlinecite{Ivanov97PRE} for the phonon-assisted kinetics  of
bulk excitons at $T_b \leq  T_c$ ($T_c$ is the critical temperature for
Bose-Einstein condensation of excitons in bulk semiconductors),
similar to this case  it depends
only on  two control  parameters of  the system,  $T_0$ and $T_b$.
Furthermore, the only gross information from a particular shape  of
the initial distribution  at $t=0$ is  absorbed by the  start time
$t_c$ of the adiabatic stage of thermalization.   This
is shown in the Fig.~\ref{fig1}(a), where we vary the  parameters
of the initial Gaussian  distribution.  In  turn, the first  transient
depends upon initial  distribution and lasts only a few  scattering
times  $\tau_{\rm  sc}$,  as  illustrated in Fig.~\ref{fig1}(b).

%

    Because  at  any  nonzero  bath  temperature  $T_b  >  0$  the
occupation number of the  ground-state mode $N^0_{\bf k_{\|}=0}  =
\exp(T_0/T_b) - 1$ is finite, the adiabatic stage ends up with 
linearized  kinetics.    The  latter  kinetics refers to the last,
exponential stage of thermalization at $t \rightarrow \infty$  and
is  characterized  by  $N_{\bf  k_{\|}=0}(t)  - N^0_{\bf k_{\|}=0}
\propto   T(t)   -   T_b   \propto   e^{-\lambda_0   t}$.      For
statistically-degenerate  QW  excitons  the  thermalization   time
$\tau_{\rm th} = 1/\lambda_0(T_b,T_0)$ reaches its smallest values
at bath temperatures  $E_0/k_B \ll T_b  \lesssim T_0$.   While the
above inequality  does not  usually hold  in the  experiments with
statistically-degenerate indirect excitons in GaAs/AlGaAs  coupled
QWs,  \cite{Butov94,Butov98,Butov99,Butov01,Larionov2001,Krivolapchuk2001} 
we  describe  a  special  design  of GaAs-based
microcavities  (MCs)  for  optimization  of the LA-phonon-assisted
thermalization  kinetics  of  low-density  MC  polaritons.  In the
proposed microcavities with a large positive detuning between  the
cavity  and  QW  exciton  modes,  the MC polaritons have radiative
lifetimes  on  a  100 ps  -   1 ns  time  scale,  so  that   the
phonon-assisted  relaxation  towards  well-developed Bose-Einstein
statistics  with  large   occupation  numbers  can   optically  be
visualized.   Thus, the  MC design  we discuss  is an  interesting
alternative to the semiconductor microcavities with zero  detuning
between the cavity mode  and QW excitons, where  huge nonclassical
occupation  numbers  of  the  low-energy  MC polariton states have
recently   been   observed.
\cite{Savvidis00,Tartakovskii00,Boeuf00,Stevenson00,Baumberg00,Savvidis01}

    In numerical evaluations  we use $M_x  = 0.21m_0$, $v_s  = 3.7
\times 10^5$ cm/s,  and $D_x =  15.5$ eV, relevant  to GaAs/AlGaAs
coupled QWs, and $M_x = 0.86  m_0$, $v_s = 4.1 \times 10^5$  cm/s,
and $D_x  = 6.9$  eV, relevant  to single  ZnSe QWs,  respectively
($m_0$ is the free electron mass).  Note that the disorder-induced
scattering and localization processes, not included in our  model,
are relatively strong in  up-to-date ZnSe-based QWs and  require a
separate analysis.  \cite{Takagahara85,Baranovskii98}
However, in the very recent work \cite{Bradford01} the first fabrication
of high-quality MgS/ZnSe/MgS QWs with less than $1$ ML fluctuations of the
well width and, therefore, with extremely low inhomogeneous broadening
has been reported.

    In  Sec.~\ref{sec2},  the  Boltzmann  equation  is adapted in
order  to  formulate  the  acoustic-phonon-assisted  kinetics   of
statistically-degenerate  QW  excitons.    We  also  discuss  some
approximations  for  the  form-factor  $F_z(\chi)$,
which  describes  the
relaxation of momentum conservation  in scattering of QW  excitons
by bulk LA-phonons.

    In  Sec.~\ref{sec3},  we  find  the  generic  solution of the
acoustic-phonon-assisted kinetics  from a  strongly nonequilibrium
initial  distribution  of  QW  excitons   $N_{\varepsilon}(t=0)$
towards a  final Bose-Einstein  distribution $N_{\varepsilon}^{0}$
with   large   occupation   numbers   of  the  low-energy  states,
$N_{\varepsilon \simeq 0}^{0} \gtrsim 1$, where
$\varepsilon$ is the dimensionless energy defined by 
$\varepsilon= \hbar^2 {\textbf k^2_{\|}}/(2 M_x E_0)$.  
The generic solution is
independent of a particular shape of the initial distribution  and
describes the adiabatic stage  of thermalization, which starts  at
$t=t_c$ in a few characteristic scattering times after $t=0$.   We
show that at $t \gg t_c$ thermalization of high-energy QW excitons
is given  by 
$\delta T(t)  \propto e^{-  \lambda_0 t}/t$,  where $\lambda_0$ is
the  lowest  positive   eigenvalue  of  the   relevant  linearized
kinetics.

    In Sec.~\ref{sec4},  the linearized  phonon-assisted kinetics
of  statistically-degenerate   QW  excitons   is  formulated   and
analyzed.   We show  that the  eigenvalues $\{  \lambda \}$ of the
linear collision  integral form  a continuous  spectrum, $\infty >
\lambda  \geq  \lambda_0  >  0$,  separated from the nondegenerate
eigenvalue   $\lambda   =   0$,   and   that   the   corresponding
eigenfunctions  $\{  \psi_{\varepsilon}(\lambda)  \}$  have  three
well-defined  isolated  critical  points.    The dependence of the
marginal eigenvalue $\lambda_0$ on the bath ($T_b$) and degeneracy
($T_0$) temperatures is studied.

    In Sec.~\ref{sec5}, straightforward numerical simulations  of
the phonon-assisted relaxation kinetics at $T_b \lesssim T_0$  are
compared with the generic solution of the 
quantum Boltzmann equation.   We
also show that at the  beginning of the adiabatic stage  of
thermalization,  at  $t_c  \leq  t  \lesssim  \lambda_0^{-1}$, the
population  dynamics  of  the  ground-state  mode  is given by 
$N_{\varepsilon=0}(t)  \propto  (1+\chi t)^{\nu}$, where the
parameters $\chi$  and $\nu$ are calculated analytically.   
Furthermore, we
propose  a  particular  design  of GaAs-based microcavities (MCs),
which compromises the efficiency of LA-phonon-assisted  scattering
(the  density  of  states  is  $\propto  M_x$) with the degeneracy
temperature $T_0 \propto M_x^{-1}$,  i.e., allows us to  avoid the
bottleneck effect  in relaxation  and, therefore,  to optimize the
thermalization kinetics of low-density QW excitons.

In the Appendix, some relationships relevant to the thermalization
dynamics of quasi-equilibrated high-energy QW excitons are given.

\section{BOLTZMANN KINETIC EQUATION FOR DEGENERATE QW EXCITONS}
\label{sec2}

    For  hot  QW   excitons,  which  are   in-plane  isotropically
distributed  at  $t=0$,  the  thermalization  kinetics due to bulk
LA-phonons can  be treated  in one-dimensional  energy space  [see
Eq.~(2) of Ref.~\onlinecite{Ivanov99PRB}].  In the following  we
express energy  $E$ and  temperature $T$  (as well  as $T_b$  and
$T_0$) in terms of $E_{0}$,  i.e., we use the  dimensionless
values of $E \to \varepsilon=E/E_{0}$, and $T \to (k_B T)/E_{0}$.  In
order to derive and analyze the generic solution for relaxation at
$T_b <  T_0$, it  is convenient  to rewrite  the above equation in
terms of the variable
\begin{equation}
f_{\varepsilon}(t) = \frac{N_{\varepsilon}(t)-N^{0}_{\varepsilon}}
{T_b (N^{0}_{\varepsilon})'} ,
\label{2.1}
\end{equation}
    where   $N_{\varepsilon}(t)$   and   $N^{0}_{\varepsilon}    =
1/[e^{(\varepsilon-\mu)/T_b}-1]$  are   the  current   and  (final)
equilibrium distribution functions  of QW excitons,  respectively,
and  the  chemical  potential  $\mu$  is  given  by  $\mu  =   T_b
\ln(1-e^{-T_0/T_b})$.  In this  case the kinetic equation  reduces
to
\begin{widetext}
\begin{equation}
\frac{\partial}{\partial t}f_{\varepsilon}(t) =
-\frac{4}{\tau_{\rm sc}}\left[ \int^{\theta_{\rm S}(\varepsilon)}_{0}
F_{\rm S}(\varepsilon, \varepsilon_{1})
{\mathcal L}_{\rm S}(\varepsilon, \varepsilon_{1},t)
\varepsilon_{1} d \varepsilon_{1} 
+\int^{\infty}_{\theta_{\rm AS}(\varepsilon)}
F_{\rm AS}(\varepsilon, \varepsilon_{1})
{\mathcal L}_{\rm AS}(\varepsilon, \varepsilon_{1},t) \varepsilon_{1}
d \varepsilon_{1} \right] ,
\label{2.2}
\end{equation}
\end{widetext}
    where the Stokes (S) and anti-Stokes (AS) collision integrands
are
\begin{subequations}
\begin{eqnarray}
{\mathcal L}_{\rm S}(\varepsilon, \varepsilon_{1},t) &=& \left[
f_{\varepsilon}(t) - f_{\varepsilon - \varepsilon_{1}} (t) \right]
\left( 1+n^{\rm ph}_{\varepsilon_{1}}+N^{0}_{\varepsilon -
\varepsilon_{1}} \right)
\nonumber \\
&& + \ T_b (N^{0}_{\varepsilon - \varepsilon_{1}})' f_{\varepsilon}(t)
f_{\varepsilon - \varepsilon_{1}}(t)  ,
\label{2.3a} \\
{\mathcal L}_{\rm AS}(\varepsilon, \varepsilon_{1},t) &=& \left[
f_{\varepsilon}(t) - f_{\varepsilon + \varepsilon_{1}} (t) \right]
\left( n^{\rm ph}_{\varepsilon_{1}} - N^{0}_{\varepsilon +
\varepsilon_{1}} \right)
\nonumber \\
&& - \ T_b (N^{0}_{\varepsilon + \varepsilon_{1}})' f_{\varepsilon}(t)
f_{\varepsilon + \varepsilon_{1}}(t) .
\label{2.3b}
\end{eqnarray}
\label{2.3}
\end{subequations}
    The  distribution  of  thermal  bulk  phonons  is given by the
Planck    formula,     i.e.,    $n^{\rm     ph}_{\varepsilon}    =
1/(e^{\varepsilon/T_b}- 1)$.   The  scattering time  is defined by
$\tau_{\rm sc}  = (\pi^2  \hbar^4 \rho)/(D^2_x  M_x^3 v_s)$.   The
functions  $\theta_{\rm  S/AS}(\varepsilon)$,  which determine the
integration   limits   on   the   right-hand-side of
Eq.~(\ref{2.2}), are
\begin{subequations}
\begin{eqnarray}
\theta_{\rm S}(\varepsilon) &=& \left\{
\begin{array}{ll}
0, & \varepsilon \le 1/4 \\
2\sqrt{\varepsilon} - 1,  & 1/4 <  \varepsilon \le 1 \\
\varepsilon, & 1 <  \varepsilon ,  \\
\end{array}
\right.
\label{2.4a}
\\
\theta_{\rm AS}(\varepsilon) &=& \left\{
\begin{array}{ll}
1-2\sqrt{\varepsilon},  & \varepsilon \le 1/4 \\
0, & \varepsilon > 1/4 . \\
\end{array}
\right.
\label{2.4b}
\end{eqnarray}
\label{2.4}
\end{subequations}

The functions $F_{\rm S/AS}(\varepsilon, \varepsilon_{1})$
in Eq.~(\ref{2.2}) are given by
\begin{widetext}
\begin{eqnarray}
F_{\rm S/AS}(\varepsilon, \varepsilon_{1}) =
\int^{u^{\rm S/AS}_{\varepsilon,
\varepsilon_{1}}}_{d^{\rm S/AS}_{\varepsilon, \varepsilon_{1}}}
{\left|F_z(a \varepsilon_{1} \alpha)\right|}^2
\left\{
\left[ \left( u^{\rm S/AS}_{\varepsilon, \varepsilon_{1}} \right)^2
-\alpha^2 \right]
\left[\alpha^2 -
\left(d^{\rm S/AS}_{\varepsilon, \varepsilon_{1}}\right)^2 \right]
\right\}^{-1/2}
d \alpha  ,
\label{2.5}
\end{eqnarray}
\end{widetext}
    where $u^{\rm S/AS}_{\varepsilon, \varepsilon_{1}} = \left[  1
- (\sqrt{\varepsilon} - \sqrt{\varepsilon \mp  \varepsilon_{1}})^2
/ \varepsilon_{1}^2 \right]^{1/2}$ and $d^{\rm S/AS}_{\varepsilon,
\varepsilon_{1}}   =    \left[   1    -   (\sqrt{\varepsilon}    +
\sqrt{\varepsilon  \mp  \varepsilon_{1}})^2  /   \varepsilon_{1}^2
\right]^{1/2}$. 
The dimensionless parameter $a$ in the argument of the form-factor
function $F_z$  is given by $a = (L_z M_x v_s)/\hbar$.

    In order to estimate the contribution of the form-factor $F_z(\chi)$
to    the    spectral    functions    $F_{\rm   S/AS}(\varepsilon,
\varepsilon_{1})$ one can analyze Eq.~(\ref{2.5}) for $\varepsilon
\to  0$.    In  this  case  only  the  spectral  function  $F_{\rm
AS}(\varepsilon,  \varepsilon_{1})$  is  relevant  to  the kinetic
Eq.~(\ref{2.2}), and Eq.~(\ref{2.5}) yields
\begin{equation}
F_{\rm AS}(0, \varepsilon_{1}) =
\frac{\pi}{2} \sqrt{\frac{ \varepsilon_{1}}{ \varepsilon_{1} -1}} \
{\left|F_z \left[a \sqrt{ \varepsilon_{1}( \varepsilon_{1} -1)} \right]
\right|}^2 .
\label{2.6}
\end{equation}
    For  $\varepsilon_{1}  \gg  1$,  the  spectral  width  $\Delta
\varepsilon$ (FWHM) of  $F_{\rm AS}(0, \varepsilon_{1})$  is given
by $\Delta\varepsilon \simeq 2.26/a$ and determined  solely
by $F_z(a \varepsilon_1)$.  For thickness $L_z=8$ nm of GaAs  QWs
(see Ref.~\onlinecite{Butov01}) one  gets $a  \simeq 0.054$ and,
therefore, $\Delta  \varepsilon \simeq  42$.   The latter estimate
can be rewritten  in the dimensional  energy units as  $\Delta E =
E_0 \Delta  \varepsilon =  4.52 \hbar  v_s/L_z \simeq  1.44$ meV.
Thus the spectral band  of LA-phonons, which scatter  a low-energy
QW exciton, can  be evaluated as  $\Delta E \sim  v_s \Delta p_z$,
where  $\Delta  p_z  \sim  \hbar/L_z$  is  the  uncertainty of the
momentum in the $z$-direction  due to the QW  spatial confinement.
The above estimates show that for relatively cold QW excitons with
energies  $\varepsilon  \leq  \Delta  \varepsilon$  (the effective
temperature $T^{\rm eff} = \Delta E/k_B \simeq 16.5$ K for  $L_z
=  8$ nm)   the  form-factor   can  be   approximated  by  $F_z(a
\varepsilon_{1} \alpha) = F_z(0) = 1$.  In this case the  integral
on the r.h.s. of Eq.~(\ref{2.5}) can be calculated explicitly:
\begin{equation}
F_{\rm S/AS}(\varepsilon, \varepsilon_{1}) =
\frac{1}{d^{\rm S/AS}_{\varepsilon, \varepsilon_{1}}} \
\Phi\left[{\left(
\frac{d^{\rm S/AS}_{\varepsilon, \varepsilon_{1}}}
{u^{\rm S/AS}_{\varepsilon, \varepsilon_{1}}} \right)}^2 \right] ,
\label{2.7}
\end{equation}
where
\begin{equation}
\Phi(\xi) = \left\{
\begin{array}{ll}
- i \left\{
F\left[\arcsin\left(\sqrt{\xi}\right),\frac{1}{\xi}\right] -
K\left( \frac{1}{\xi} \right) \right\},  & 0 \le \xi < 1 \\
K\left( \frac{1}{\xi} \right), & \xi < 0 . \\
\end{array}
\right.
\label{2.8}
\end{equation}
    Here,   $F(\phi,m)=\int_{0}^{\phi}\left[1-   m  \sin^2(\theta)
\right]^{-1/2} d  \theta$ and  $K(m)=F(\pi/2,m)$ are  the elliptic
and  the   complete  elliptic   integrals  of   the  first   kind,
respectively.

\section{ GENERIC SOLUTION }
\label{sec3}

\begin{figure}
\includegraphics{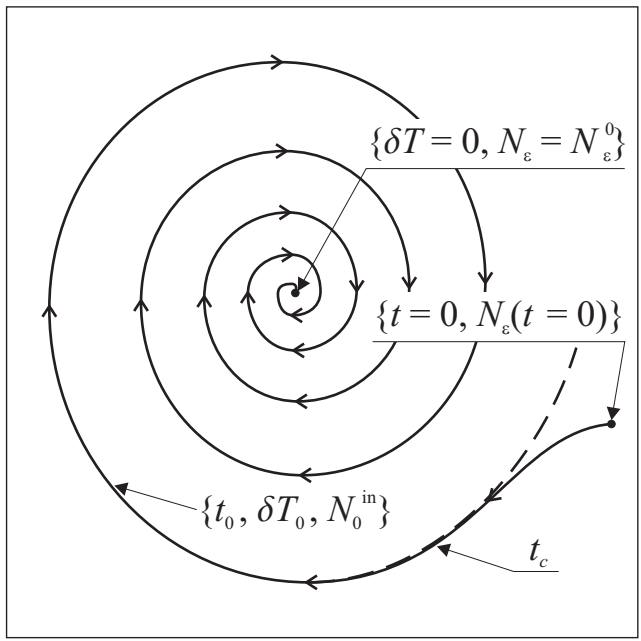}
\caption[Figure 2]{
    Schematic  picture  of  the  relaxation  kinetics at $T_b\lesssim T_0$. 
The reference (start) point $\{t_0, \delta T_0, N_{0}^{\rm
in}  \}$  unambiguously  determines  the  calculated  (solid line)
evolution from the initial noequilibrium distribution $N_{\varepsilon}(t=0)$.
The dashed  line corresponds to the  generic solution.
$t_c$ is  the start point  of adiabatic  stage of  evolution.   For the
times $t - t_c \ge \tau_{\rm sc}$ both  evolution spirals almost
coincide showing a unique path towards the final equilibrium 
distribution $\{\delta  T = 0, N_{\varepsilon} = N_{\varepsilon}^{0} \}$.
}
\label{fig2}
\end{figure}

In this Section the thermalization kinetics at $T_b \le T_0$ is described in
terms of a generic solution, which weakly correlates with the initial
distribution $N_{\varepsilon}(t=0)$. In particular, we  
derive a thermalization  law  for  high-energy excitons and examine
nonequilibrium distribution  of low-energy  QW excitons.   We also
specify a reference point for the generic solution, i.e., a  set
of  parameters,  which  unambiguously  determines  the  calculated
evolution.
Schematic picture of the relaxation kinetics in  phase
space is shown in  Fig.~\ref{fig2}.  The generic  solution
is relevant to  the times $t >  t_c$,
where $t_c$  is the  start time of  the adiabatic stage  of evolution.

%

As was emphasized in Sec.~\ref{Introduction}, 
the generic  solution  assumes a
quasi-equilibrium distribution of  \textit{high-energy}    excitons
($\varepsilon > 1/4$):      $N_{\varepsilon   >  1/4}(t)  =
1/[e^{(\varepsilon-\tilde\mu(t))/T(t)}-1]$.                   This
quasi-equilibrium  distribution  is  characterized  by  the effective
time-dependent  temperature  $T(t)=T_b+\delta  T(t)$  and chemical
potential $\tilde\mu(t)$ whose time variations are supposed to be
small,  so  that  $\tilde\mu(t)$  can  be  taken  the  same as the
chemical    potential    of    final    equilibrium   distribution
$\tilde\mu(t)\simeq\mu$.  Indeed, for $T(t) < T_0$ the chemical
potential is given by 
$\tilde\mu(t) \simeq -T(t) e^{-T_0/T(t)} \ll T(t)$ and,
therefore, $|\delta \mu|= |\tilde\mu(t) - \mu| \ll \delta  T(t)$. 
Since at the final stage of relaxation  kinetics
the effective temperature  of high-energy excitons  approaches the
bath temperature ($\delta T(t) \to 0$), starting from some  moment
in time  one meets  the condition  $\delta T(t)  \ll T_b$ (provided that
$T_b >  0$).   Therefore we can linearize  the 
quasi-equilibrium distribution function with respect to $\delta  T(t)$.  
In this  case one obtains
\begin{equation}
f_{\varepsilon > 1/4}(t) = -[\delta T(t)/T_b^2]\varepsilon .
\label{3.1.1}
\end{equation}

On the  other hand,  as we  show below,  
a \textit{low-energy} kernel ($0< \varepsilon \leq 1/4$)  
of the distribution  function characterizes the non-equilibrium QW excitons.
With increasing time  $t \ge t_c$, the low-energy kernel shrinks in
energy space, and the ratio   
$|N_{\varepsilon  \leq  1/4}  -  N_{\varepsilon  \leq  1/4}^0|/
N_{\varepsilon   \leq   1/4}^0$ 
decreases approaching the limit when the linearized kinetics
becomes valid.

\subsection{ Nonequilibrium distribution of low-energy QW excitons }
\label{sec3.2}

    In order to  analyze the evolution  of low-energy QW  excitons
within  the  scenario  described  in  the  previous subsection, we
substitute Eq.~(\ref{3.1.1})  into Eq.~(\ref{2.2})  and get  the
reduced kinetic equation for $\varepsilon \leq 1/4$:
\begin{equation}
\frac{\partial}{\partial t} f_{\varepsilon}(t) =
- \left[\xi_0(\varepsilon) + \xi_1(\varepsilon)\delta T(t)\right]
f_{\varepsilon}(t) + \eta(\varepsilon) \delta T(t) ,
\label{3.2.1}
\end{equation}
where
\begin{equation}
\xi_0(\varepsilon)=
\frac{4}{\tau_{\rm sc}}
\int^{\infty}_{1- 2 \sqrt{\varepsilon}}
F_{\rm AS}(\varepsilon, \varepsilon_{1})
\left(
n^{\rm ph}_{\varepsilon_{1}} - N^{0}_{\varepsilon + \varepsilon_{1}}
\right) \varepsilon_{1} d \varepsilon_{1} ,
\label{3.2.2}
\end{equation}
\begin{equation}
\xi_1(\varepsilon)=
\frac{4}{\tau_{\rm sc}T_b}
\int^{\infty}_{1- 2 \sqrt{\varepsilon}}
F_{\rm AS}(\varepsilon, \varepsilon_{1})
\left(N^{0}_{\varepsilon + \varepsilon_{1}} \right)'
\varepsilon_{1} (\varepsilon+\varepsilon_{1})
d \varepsilon_{1} ,
\label{3.2.3}
\end{equation}
\begin{eqnarray}
\eta(\varepsilon)=
- \frac{4}{\tau_{\rm sc}T_b^2}
\int^{\infty}_{1- 2 \sqrt{\varepsilon}}&&
F_{\rm AS}(\varepsilon, \varepsilon_{1})
 \left(
n^{\rm ph}_{\varepsilon_{1}} - N^{0}_{\varepsilon + \varepsilon_{1}}
\right)
\nonumber\\
&&\times
\varepsilon_{1} (\varepsilon+\varepsilon_{1})
d \varepsilon_{1} ,
\label{3.2.4}
\end{eqnarray}
    and  $F_{\rm  AS}(\varepsilon+\varepsilon_{1})$  
is  given  by  Eq.~(\ref{2.7}).  Equation
(\ref{3.2.1}),  which  describes  the  population  dynamics of the
low-energy states during the adiabatic stage of thermalization ($t
\geq t_c$),  is a  linear inhomogeneous  differential equation for
$f_{\varepsilon \leq 1/4}$.  Its complete solution can be  written
as a sum of the homogeneous and inhomogeneous contributions:
\begin{eqnarray}
f_{\varepsilon \leq 1/4}(t)=
f_{\varepsilon \leq 1/4}^{\rm hom}(t)
+\eta(\varepsilon) \int_{t_0}^{t}&&
e^{-(t-\tau)\xi_0(\varepsilon)-\rho(t,\tau)\xi_1(\varepsilon)}
\nonumber\\
&&\times
\delta T(\tau) d \tau,
\label{3.2.5}
\end{eqnarray}
where
\begin{equation}
f_{\varepsilon \leq 1/4}^{\rm hom}(t)=
e^{-(t-t_0)\xi_0(\varepsilon)-\rho(t,t_0)\xi_1(\varepsilon)}
f_{\varepsilon \leq 1/4}(t=t_0),
\label{3.2.6}
\end{equation}
\begin{equation}
\rho(t,t_1)=
\int_{t_1}^{t} \delta T(\tau) d \tau ,
\label{3.2.7}
\end{equation}
    and  $t_0$  is  an  arbitrary  reference  (start) time for the
calculated evolution, i.e., $t_0 \geq t_c$ (see Fig.~\ref{fig2}).

    Thus the adiabatic stage  of the relaxation kinetics  into the
lower-energy  states  is  completely
determined by $N_{\varepsilon}(t)  =  N^0_{\varepsilon} +
T_b(N^0_{\varepsilon})'   f_{\varepsilon}(t)$,
provided   that  one   knows  the   reference  (start)
distribution    $f_{\varepsilon     \leq    1/4}(t=t_0)$     and
thermalization law $\delta  T = \delta  T (t)$ for  high-energy QW
excitons.      Note   that   in   a   sharp   contrast   with  the
acoustic-phonon-assisted relaxation kinetics  at $T_b \leq  T_c$ in
three-dimensional systems,  \cite{Ivanov97PRE} the  thermalization
dynamics of  low-energy QW  excitons depends  upon the homogeneous
contribution     $f_{\varepsilon \leq 1/4}^{\rm hom}$ given  by
Eq.~(\ref{3.2.6}).

    As will be  shown in the  next subsection, the  thermalization
law for  quasi-equilibrium high-energy  QW excitons  ($\varepsilon
> 1/4$) is given by
\begin{equation}
\delta T(t) =
\left(
\frac{ \delta T_0}{\lambda_1 - \lambda_0}
\right)
\frac{  e^{-\lambda_0 (t - t_0)} -  e^{-\lambda_1 (t - t_0)}  }
{t-t_0} ,
\label{3.2.8}
\end{equation}
    where   $\lambda_0    =   \xi_0(0)$    characterizes  the inverse
thermalization   time at $t \to \infty$,   
$\lambda_1   =   \xi_0(1/4)$  is  another
characteristic parameter  relevant to  the beginning  of the adiabatic
stage  ($\lambda_1  \gg  \lambda_0$),  and  $\delta  T_0  = \delta
T(t=t_0)$  determines  reference  (start)  effective   temperature
$T(t=t_0)  =  T_b  +  \delta  T_0$  for  the calculated evolution.
Equation  (\ref{3.2.8})  is  valid  for  $t  \geq t_0$.  Using the
thermalization  law  (\ref{3.2.8})  we  find  the  integral on the
r.h.s. of Eq.~(\ref{3.2.7}):
\begin{eqnarray}
\rho(t,t_1) &=&
\frac{ \delta T_0}{\lambda_1 - \lambda_0}
 \left\{
Ei[-\lambda_0 (t - t_0)] - Ei[-\lambda_0 (t_1 - t_0)] \right.
\nonumber\\
&& \left.
-Ei[-\lambda_1 (t - t_0)] + Ei[-\lambda_1 (t_1-t_0)]
 \right\},
\label{3.2.9}
\end{eqnarray}
    where  $Ei(z)=  -\int_{-z}^{\infty}  (e^{-t}/t)  dt$  is   the
exponential integral function.

    The reference distribution $f_{\varepsilon \leq  1/4}(t=t_0)$,
which  determines  the  homogeneous  solution  (\ref{3.2.6}),   is
independent  of  the  initial  distribution  of  hot  QW excitons,
$N_{\varepsilon}(t=0)$.      There   is,   however,   an  integral
relationship  between  $\delta   T_0$  and  $f_{\varepsilon   \leq
1/4}(t=t_0)$,  which  we  will  discuss in subsection \ref{sec3.4}
along  with  possible  approximations  for  $f_{\varepsilon   \leq
1/4}(t=t_0)$.

\subsection{Thermalization of high-energy QW excitons}
\label{sec3.3}

    The temperature change $\delta T = \delta T(t)$  characterizes
the time evolution  of high-energy quasi-equilibrated particles 
through Eqs.~(\ref{2.1}) and (\ref{3.1.1}).  In order to derive  the
temperature law (\ref{3.2.8})  we substitute $f_{\varepsilon  \leq
1/4}$ and $f_{\varepsilon > 1/4}$ given by Eq.~(\ref{3.2.5})  and
(\ref{3.1.1}),   respectively,    into   the    kinetic   equation
(\ref{2.2}):
\begin{equation}
\frac{\partial}{\partial t} \delta T(t) =
- \left[\alpha_0 + \alpha_1(t)
\right] \delta T(t) + \beta \delta T^{2}(t) + \gamma(t)  ,
\label{3.3.1}
\end{equation}
where
\begin{eqnarray}
\gamma (t)= - \frac{4 T^{2}_{b}}{\tau_{\rm sc}\varepsilon_c}
\int^{1/4}_{0} &&
F_{\rm S}(\varepsilon_c, \varepsilon_c -  \varepsilon)
\left( 1 + n^{\rm ph}_{\varepsilon_c -\varepsilon }
N^{0}_{\varepsilon} \right)
\nonumber \\
&&\times  f_{ \varepsilon}(t)
(\varepsilon_c - \varepsilon ) d \varepsilon ,
\label{3.3.2}
\end{eqnarray}
    and  the  parameters   $\alpha_0$,  $\beta$,  and   functional
$\alpha_{1}(t)  =  \alpha_{1}[f_{\varepsilon  \leq  1/4}(t)]$  are
defined in Appendix.  Equations (\ref{3.3.1})-(\ref{3.3.2})  refer
to  some  energy  $\varepsilon_c  \gtrsim  1$ from the high-energy
domain $\varepsilon > 1/4$.

    In the adiabatic stage of relaxation, when $t > t_c$, one  has
$\delta T(t)/T_b \ll 1$ so that $\alpha_1$ and $\beta \delta  T^2$
can  be  neglected  on  the  r.h.s. of Eq.~(\ref{3.3.1}), because
$\left|\alpha_1(t)  \right|  \ll  \alpha_0$  and  $|\beta   \delta
T^2(t)| \ll |\gamma(t)|$.  At the end of the adiabatic stage ($t -
t_c    \gg    \lambda_0^{-1}$)    the    distribution     function
$N_{\varepsilon}(t)$ is already very close to  $N_{\varepsilon}^0$
even for small energies $\varepsilon \leq 1/4$ and, therefore, the
phonon-assisted  relaxation  kinetics  becomes  exponential, i.e.,
$\delta     T(t)     \propto     e^{-\lambda_0     t}$.        For
statistically-degenerate QW excitons ($N_{\varepsilon=0}^0 \gg 1$)
one  estimates  from   Eqs.~(\ref{3.2.2})  and  (\ref{b1})   that
positive  $\lambda_0  =  \xi(0)$   is  much  less  than   positive
$\alpha_0$.    In  this  case  Eq.~(\ref{3.3.1})  can  be  solved
iteratively.  The first iteration, which can formally be  obtained
by   neglecting   the   time   derivative   on   the   r.h.s.   of
Eq.~(\ref{3.3.1}), yields
\begin{equation}
\delta T(t) = \gamma(t) / \alpha_0 .
\label{3.3.3}
\end{equation}
    For  the  same  time  domain  $t  -  t_c  \gg  \lambda_0^{-1}$
Eq.~(\ref{3.2.5}) yields the following approximation:
\begin{equation}
f_{\varepsilon \leq 1/4}(t) = e^{-(t-t_0)\xi_0(\varepsilon)}
f_{\varepsilon}(t=t_0) +
\frac{\eta(\varepsilon)}{\xi_0(\varepsilon)} \delta T(t) .
\label{3.3.4}
\end{equation}
    By  substituting   Eq.~(\ref{3.3.4})  into   the  r.h.s.   of
Eq.~(\ref{3.3.2}) one derives from Eq.~(\ref{3.3.3}):
\begin{eqnarray}
\delta T(t) = \frac{1}{\tilde \alpha}
\int^{1/4}_{0}&&
F_{\rm S}(\varepsilon_c, \varepsilon_c -  \varepsilon)
\left( 1 + n^{\rm ph}_{\varepsilon_c -\varepsilon }
N^{0}_{\varepsilon} \right) f_{ \varepsilon}(t_0)
\nonumber \\
&&\times  
e^{-(t - t_0) \xi_0(\varepsilon)}
(\varepsilon_c - \varepsilon ) d \varepsilon,
\label{3.3.6}
\end{eqnarray}
    where the  constant ${\tilde\alpha}$  is defined  in Appendix.
We  can  further  simplify  Eq.~(\ref{3.3.6}) taking into account
that $\xi_0(\varepsilon)$ is a monotonously increasing function of
energy and that at large $t$ only small vicinity of $\varepsilon =
0$  contributes  to  the  integral.    Finally, we end up with the
asymptotic law:
\begin{equation}
\delta T (t) = \left(
\frac{
{\tilde\gamma}_0 }{ {\tilde \alpha}
}
\right) 
\frac{e^{- \xi_0(0) (t-t_0)} - e^{-\xi_0(1/4) (t-t_0)}}{t - t_0},
\label{3.3.7}
\end{equation}
    where the  constant ${\tilde\gamma}_0$  is given  in Appendix.
Equation (\ref{3.3.7}) is identical to Eq.~(\ref{3.2.8}) provided
that the  reference (start)  temperature $T_b+\delta  T(t=t_0)$ of
the high-energy quasi-equilibrated QW excitons is determined by
\begin{equation}
\delta T_0 \equiv
\delta T(t=t_0) = ( {\tilde\gamma}_0 / {\tilde \alpha} )
(\lambda_1 - \lambda_0).
\label{3.3.8}
\end{equation}

\begin{figure}
\includegraphics{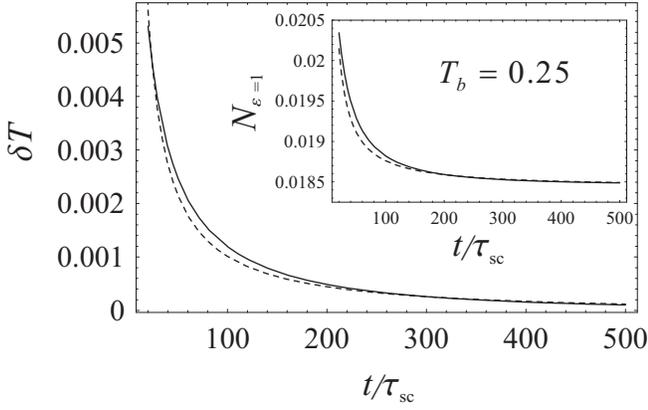}
\caption[Figure 3]{
    Time dependence of $\delta  T$ and $N_{\varepsilon =  1}$ 
(inset) calculated  by using  the thermalization  law (\ref{3.2.8})
(dashed lines) and  by direct numerical  modelling of the  initial
kinetic Eq.~(\ref{2.2}) (solid lines).  The control parameters  $T_b$
and $T_0$ are the same as in Fig.~\ref{fig1}.
}
\label{fig3}
\end{figure}

    While         the         above         derivation          of
Eqs.~(\ref{3.3.3})-(\ref{3.3.8}) assumes that $t  - t_c \geq t  -
t_0 \gg \lambda_0^{-1}$, we have checked numerically that \textit{the
temperature  law}  (\ref{3.2.8})  \textit{holds  through  the  whole
adiabatic stage}, i.e., for  the time interval $0  \leq t - t_0  <
\infty$.    Furthermore,  the  numerical  evaluations also clearly
indicate  that  the  start  temperature  $\delta  T(t  =  t_0)$ is
practically independent of the local energy $\varepsilon_c \gtrsim
1$ used  in the  derivation of  Eqs.~(\ref{3.3.1})-(\ref{3.3.8}).
In  Fig.~\ref{fig3}  we  plot  $\delta  T  =  \delta  T(t)$   and
$N_{\varepsilon = 1} = N_{\varepsilon = 1}(t)$ calculated by using
the thermalization law (\ref{3.2.8}) (dashed lines) and by  direct
numerical modelling of the initial kinetic Eq.~(\ref{2.2}) (solid
lines), respectively.

%

    For the  time domain  $ \lambda_1^{-1}  \leq t  - t_0 \lesssim
\lambda_0^{-1}$  the  thermalization  law  (\ref{3.2.8})  can   be
approximated by $\delta  T(t) = (\lambda_1-  \lambda_0)^{-1}\delta
T_0  /(t-t_0)$.    The  latter  dependence  $\delta  T(t)  \propto
1/(t-t_0)$ is consistent with  that found for the  adiabatic stage
of the phonon-assisted relaxation kinetics of 3D bosons (excitons)
at  $T_b  \leq  T_c$,  when  Bose-Einstein  condensate  builds up.
\cite{Ivanov97PRE} For 2D systems at nonzero $T_b$ the  occupation
number of  the ground-state  mode $N_{\varepsilon=0}^0$  is
always final.   Therefore  the exponential  kinetics $\delta  T(t)
\propto   e^{-   \lambda_0    (t-t_0)}$,   which   results    from
Eq.~(\ref{3.2.8}) for $t -  t_0 \gtrsim \lambda_0^{-1}$, develops  at
the       final       stage       of       relaxation,        when
$|N_{\varepsilon=0}(t)-N_{\varepsilon=0}^0|/N_{\varepsilon=0}^0
\ll  1$.    Note  that  because  for  statistically-degenerate  QW
excitons, when $T_b \lesssim T_0$ and $N_{\varepsilon=0}^0 \gg 1$,
one has $\lambda_0^{-1}  \gg \tau_{sc}$ (see  Section \ref{sec4}),
the two modes of behavior,  $\delta T \propto 1/t$ and  $\delta T
\propto e^{- \lambda_0 t}$, are well-separated in time.

\subsection{Reference point for the generic solution}
\label{sec3.4}

    In  order   to  determine   the relaxation kinetics of low-energy 
particles, i.e., $N_{\varepsilon \leq 1/4} = 
N_{\varepsilon \leq 1/4}(t \geq t_c)$, one needs  to specify the reference  
distribution $f_{\varepsilon \leq 1/4}(t=t_0)$ 
and $\delta T_0 = \delta T(t=t_0) = T(t=t_0)-T_b$.  
According to Eq.~(\ref{3.2.6}), with increasing time $t-t_0$ only the 
small vicinity  of $\varepsilon = 0$  gives contribution  to the 
homogeneous  part  of the generic solution. Therefore we approximate 
$N_{\varepsilon \leq 1/4}(t=t_0) \simeq N^{\rm in}_{0} \equiv 
N_{\varepsilon=0}(t=t_0)$. The above approximation assumes that the 
reference time $t_0 \geq t_c$ is close enough to the start time $t_c$ of 
the adiabatic stage of relaxation so that $N_{\varepsilon}(t_0) \ll 
N^{0}_{\varepsilon}$ at $\varepsilon \ll 1/4$. 
Thus, using Eq.~(\ref{2.1}) we determine the 
reference distribution at $t=t_0$ by  
\begin{equation}
f_{\varepsilon \leq 1/4}(t=t_0) = 
\frac{N^{\rm in}_{0} - N^{0}_{\varepsilon}} 
{T_b (N^{0}_{\varepsilon})'}. 
\label{3.4.1}
\end{equation}
The latter expression is completely defined by the only one unknown 
parameter $N^{\rm in}_{0}$, the population of the ground-state mode 
at the reference time $t_0$. 

Equation (\ref{3.4.1}) allows us to find $\delta T_0 = \delta T(t=t_0)$  
through  the  integral relationship (\ref{3.3.6}) taken  at  $t=t_0$. 
Furthermore,  within the approximations used in the derivation of 
the asymptotic  law (\ref{3.3.7}) a  simplified (algebraic) form  of 
this  relationship,  given  by Eqs.~(\ref{3.3.8}) and (\ref{b5}), is 
valid as well. After the value of $\delta T_0$ is determined, using the 
thermalization law (\ref{3.2.8}) one can easily find the time dependence  
of the distribution function (\ref{3.1.1}) of high-energy  excitons. 
Therefore, the three parameters $t_0$, $N^{\rm in}_{0}$, and $\delta T_0$, 
completely specify the reference point for the generic solution, as 
illustrated in Fig.~\ref{fig2}.

\section{ Linearized kinetics for statistically
                   degenerate QW excitons }
\label{sec4}

    If  for  any  energy  $\varepsilon$  the distribution function
$N_{\varepsilon}$ of quantum-degenerate quasi-2D excitons is close
enough to  final $N_{\varepsilon}^0$  so that  $|N_{\varepsilon} -
N_{\varepsilon}^0|/N_{\varepsilon}^0       \ll       1$        and
$f_{\varepsilon}(t)$ becomes  small, the  phonon-assisted kinetics
can  be  linearized.    In  particular,  the  adiabatic  stage  of
thermalization at times  $t - t_0  \geq \lambda_0^{-1}$ refers  to
the linearized kinetics.  The linearized kinetics can be described
in terms of the real eigenvalues $\{ \lambda \}$ ($\lambda\geq 0$)
and       the        corresponding       eigenfunctions        $\{
\psi_{\varepsilon}(\lambda)  \}$  so  that  $f_{\varepsilon}(t)  =
\sum_{\lambda}       c_{\lambda}       \psi_{\varepsilon}(\lambda)
\exp(-\lambda t)$.   The initial kinetic  Eq.~(\ref{2.2}) reduces
to the linear Fredholm equation of the second kind with respect to
$\psi_{\varepsilon}(\lambda)$:
\begin{eqnarray}
\lambda  \psi_{\varepsilon}(\lambda) &=& \frac{4}{\tau_{\rm sc}}
\left[ \int^{\theta_{\rm S}(\varepsilon)}_{0} F_{\rm S}(\varepsilon,
\varepsilon_{1}) \tilde{\mathcal L}_{\rm S}(\varepsilon, \varepsilon_{1})
\varepsilon_{1} d \varepsilon_{1} \right.
\nonumber\\
&& + \left. \int^{\infty}_{\theta_{\rm AS}(\varepsilon)}
F_{\rm AS}(\varepsilon, \varepsilon_{1})
\tilde{\mathcal L}_{\rm AS}(\varepsilon, \varepsilon_{1})
\varepsilon_{1} d \varepsilon_{1} \right], \qquad
\label{4.1}
\end{eqnarray}
where
\begin{subequations}
\begin{equation}
\tilde{\mathcal L}_{\rm S}(\varepsilon, \varepsilon_{1}) =
\left[
\psi_{\varepsilon}(\lambda) -
\psi_{\varepsilon - \varepsilon_{1}} (\lambda)
\right]
\left(
1+n^{\rm ph}_{\varepsilon_{1}}+N^{0}_{\varepsilon - \varepsilon_{1}}
\right),
\label{4.2a}
\end{equation}
\begin{equation}
\tilde{\mathcal L}_{\rm AS}(\varepsilon, \varepsilon_{1}) =
\left[
\psi_{\varepsilon}(\lambda) -
\psi_{\varepsilon + \varepsilon_{1}} (\lambda)
\right]
\left(
n^{\rm ph}_{\varepsilon_{1}} - N^{0}_{\varepsilon + \varepsilon_{1}}
\right),
\label{4.2b}
\end{equation}
\label{4.2}
\end{subequations}
    and functions $\theta_{\rm S}$ and $\theta_{\rm AS}$ are given
by Eqs.~(\ref{2.4a}) and (\ref{2.4b}), respectively.

\begin{figure}
\includegraphics{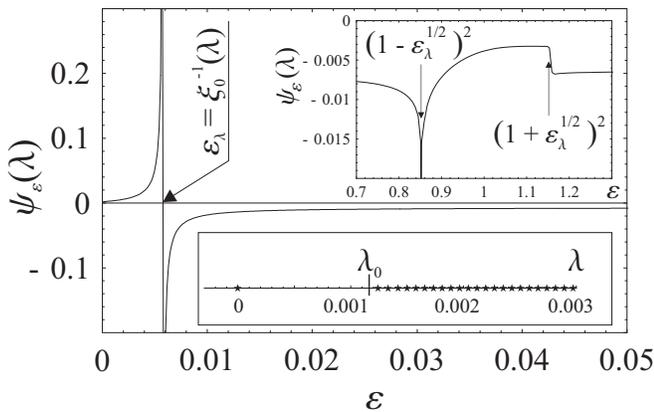}
\caption[Figure 4]{
    A        typical         shape        of     the    eigenfunction
$\psi_{\varepsilon}(\lambda)$.  The main part of the figure  shows
the  first   order  pole,   which  arises   at the  energy  region
$\varepsilon <  1/4$ ($\lambda  = 0.0043  /\tau_{\rm sc}$  in this
particular  example).    The  top  inset illustrates eigenfunction
behavior  at  critical  points  in  the high  energy band $\varepsilon
\ge1$.  The eigenvalue  spectrum ($\lambda$ in units  $1/\tau_{\rm
sc}$) is shown in the bottom inset.  The control parameters $T_b$  and
$T_0$ are the same as in Fig.~\ref{fig1}.
}
\label{fig4}
\end{figure}

    Thus  we  replace  the  solution  of  Eq.~(\ref{2.2})  by the
eigenfunction   analysis   of   the   Fredholm  integral  equation
(\ref{4.1}).  The  numerical solution of  Eq.~(\ref{4.1}) clearly
shows that  for a  given $T_b>0$  all except  one eigenvalues  $\{
\lambda  \}$  are  nondegenerate,  positive,  and  belong  to  the
continuous spectrum.  This is  illustrated in the bottom inset  of
Fig.~\ref{fig4}, where the set of eigenvalues shown by the  stars
covers  the  same  interval  $\infty  >  \lambda  \geq \lambda_0 =
\lambda_0(T_b,T_0)$ more  dense the  more discrete  points in 
energy space  are used.   The  isolated non-degenerate  eigenvalue
$\lambda=0$  is  due  to  conservation  of  the total number of QW
excitons in our model (the  only integral of motion of  the system
\cite{Cercignani88}).

%

    Since all $\{ \lambda \}$ are nondegenerate, the corresponding
eigenfunctions $\{ \psi_{\varepsilon}(\lambda) \}$ form a basis in
Hilbert energy-space.  For the  energy band $0 \leq \varepsilon  <
1/4$ one derives from Eq.~(\ref{4.1}):
\begin{equation}
\psi_{\varepsilon < 1/4}(\lambda) =
\frac{\sigma(\varepsilon)}{\lambda - \xi_0(\varepsilon)},
\label{4.3}
\end{equation}
    where $\sigma(\varepsilon)$  is a  smooth regular  function of
$\varepsilon$ given by
\begin{eqnarray}
\sigma(\varepsilon) = - \frac{4}{\tau_{\rm sc}}
\int^{\theta_{\rm S}(\varepsilon)}_{0} &&
F_{\rm S}(\varepsilon, \varepsilon_{1}) \psi_{\varepsilon +
\varepsilon_{1}}(\lambda)
\nonumber \\
&& \times \left( n^{\rm ph}_{\varepsilon_{1}} - N^{0}_{\varepsilon +
\varepsilon_{1}} \right) \varepsilon_{1} d \varepsilon_{1}, \qquad
\label{4.4}
\end{eqnarray}
    and $\xi_0(\varepsilon)$ is defined by Eq.~(\ref{3.2.2}).  Thus
the eigenfunction $\psi_{\varepsilon}  (\lambda)$ has an  isolated
singularity   (first   order   pole)   at  $\varepsilon_{\lambda}=
\xi_{0}^{-1}(\lambda)$.  The singularity is integrable in terms of
principal   value   integration.      A   typical   shape  of  the
eigenfunctions    at     $\varepsilon\!<1/4$    is     shown    in
Fig.~\ref{fig4}.    In  the energy  band  $\varepsilon  \ge  1/4$   the
eigenfunction    has    another    singularity    at   the   point
$(1-\sqrt{\varepsilon_{\lambda}})^2$.      This   singularity   is
logarithmic, i.e., integrable.  It arises when the singularity  of
$\psi_{\varepsilon  -  \varepsilon_1}(\lambda)$  at $\varepsilon -
\varepsilon_1=   \varepsilon_{\lambda}$   [see   Eq.~(\ref{4.2})]
coincides  with  the  upper  boundary of integration, $\theta_{\rm
S}(\varepsilon)$, in the  Stokes collision term  on the r.h.s.  of
Eq.~(\ref{4.1}).    The  step-like  jump  at  the  critical point
$(1+\sqrt{\varepsilon_{\lambda}})^2$  does   not  accompanied   by
discontinuity of  the eigenfunction.   The  latter critical  point
originates  from a singular  behavior  of the form-factor function
$F_{\rm  S}(\varepsilon,  \varepsilon_{1})$  at $\varepsilon_{1} =
2\sqrt{\varepsilon}  -1$.    The  features  of  the  eigenfunction
$\psi_{\varepsilon}(\lambda)$  at  points  $\varepsilon  =  (1 \pm
\sqrt{\varepsilon_{\lambda}})^2$  are  shown  in  the top inset of
Fig.~\ref{fig4}.

\begin{figure}
\includegraphics{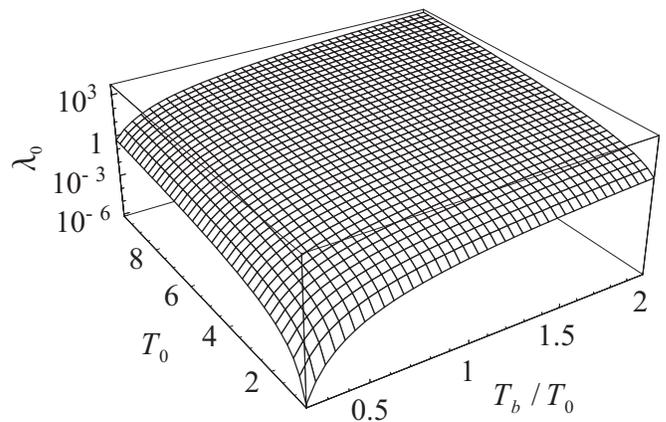}
\caption[Figure 5]{
   The inverse thermalization time $\lambda_0$ (in units $1/\tau_{\rm
sc}$) as a function of the control parameters $T_b$ and $T_0$.
}
\label{fig5}
\end{figure}

    The  marginal  point  $\lambda_0  = \lambda_0(T_b,T_0)$ of the
continuous  spectrum  of  $\{  \lambda  \}$  is  indeed  given  by
$\lambda_0 = \xi_0(0)$:   for $\lambda \rightarrow \lambda_0$  the
singularity  point  $\varepsilon_{\lambda}  \rightarrow  0$, i.e.,
approaches  its  lowest  limit.    Because $\lambda^{-1}_0$ is the
longest  relaxation  time  generated  by the continuum $\lambda_0 \leq
\lambda  <  \infty$,  the  eigenavalue  $\lambda_0$ determines the
phonon-assisted  kinetics  at  $t  \rightarrow  \infty$  and,   in
particular, yields the  characteristic thermalization time  in the
relaxation          thermodynamics          developed           in
Ref.~\onlinecite{Ivanov99PRB}.    The  dependence  of $\lambda_0$
upon the  control parameters  of the  system, $T_b$  and $T_0$, is
plotted in the Fig.~\ref{fig5}.

%

    The eigenvalue  $\lambda_0$ can  reach  both limits, i.e.,
$\lambda_0 \gg  \tau_{\rm sc}^{-1}$  and $\lambda_0  \ll \tau_{\rm
sc}^{-1}$  (see  Fig.~\ref{fig5}).     The  first   limit,  which
corresponds to the considerable acceleration of the thermalization
kinetics in comparison  with that in  the 3D case,  is due to  the
relaxation of momentum conservation in QW exciton -- bulk acoustic
phonon   scattering.      The   slowing  down  of  thermalization,
$\lambda_0\ll\tau_{\rm   sc}^{-1}$,   occurs   (i)   at  low  bath
temperatures  $T_b  \leq1$,  due  to  the exponentially decreasing
number of thermal acoustic  phonons with energy $\varepsilon  \geq
1$, and/or  (ii) for  well-developed quantum  statistics ($T_0 \gg
T_b$), due to an  effective suppression of the  stimulated kinetic
processes.    There  is  no  direct phonon-mediated interaction of
low-energy  QW  excitons  $\varepsilon  \leq  1/4$,  and  at   low
temperatures $T_b \leq  1$ the relaxation  kinetics occurs by  the
two-step process:  ``low-energy QW exciton ($\varepsilon \ll 1/4$)
$+$  phonon  ($\varepsilon_1  \simeq1$)  $\rightarrow$  QW exciton
($\varepsilon_2   =   \varepsilon   +   \varepsilon_1  \simeq  1$)
$\rightarrow$ low-energy QW exciton ($\varepsilon_4 =  \varepsilon
+   \varepsilon_1   -   \varepsilon_3   \ll   1/4$)   $+$   phonon
($\varepsilon_3 \simeq 1$)''.   The first,  anti-Stokes transition
quenches with  decreasing temperature  $T_b \leq  1$ and  yields a
temperature dependent bottleneck effect  in thermalization.   
In  turn, the critical
slowing down of  the relaxation kinetics  at $T_0 \gg  T_b$ arises
due to mutual compensation of two stimulated fluxes, into and  out
the low-energy QW  states $\varepsilon \ll  1$.  For  example, for
the ground-state  mode $\varepsilon  = 0$  the collision integrand
responsible   for   the   stimulated   kinetics   is   given    by
$N_{\varepsilon=0}(N_{\varepsilon_1\geq1}         -         n^{\rm
ph}_{\varepsilon_1\geq1})$.     At  $T_0   \gg  T_b$   the  latter
combination becomes small  in spite of  a large occupation  number
$N_{\varepsilon=0}  \gg  1$,  because  $|N_{\varepsilon_1\geq1}  -
n^{\rm              ph}_{\varepsilon_1\geq1}|               \simeq
|N^{0}_{\varepsilon_1\geq1}  -  n^{\rm   ph}_{\varepsilon_1\geq1}|
\rightarrow 0$ as a result of  a very small value of the  chemical
potential, $|\mu| \ll 1$.

    Now we can  give an alternative  prove of the  temperature law
(\ref{3.2.8})  which  refers  to  high-energy  QW  excitons   with
$\varepsilon >  1/4$.   Namely, the  function $f_{\varepsilon}(t =
t_0)$  can  be  expanded  over  the  basis  $\{ \psi_{\varepsilon}
(\lambda) \}$.  Then the solution of Eq.~(\ref{2.2}) is given by
\begin{equation}
f_{\varepsilon}(t) = \int_{\lambda_0}^{\infty}
c_{\lambda} \psi_{\varepsilon} (\lambda) e^{-\lambda (t-t_0)}
d \lambda,
\label{4.5}
\end{equation}
    where $c_{\lambda}$ are the expansion coefficients.  At  large
times  $t  -  t_0  \gg  \tau_{\rm  th}$  only  a small vicinity of
$\lambda$ near $\lambda_0$ contributes to the integral, due to the
time exponent  in the  integrand.   If now  one assumes  a regular
distribution of quasi-2D excitons  at $t = t_0$,  the coefficients
$c_{\lambda}$ smoothly depend  upon $\lambda$.   As a result,  the
approximation $c_{\lambda}  \simeq c_{\lambda_0}$  can be  used in
the integrand on the r.h.s. of Eq.~(\ref{4.5}).  Furthermore, the
eigenfunctions $\psi_{\varepsilon} (\lambda)$ have nearly the same
smooth shape at high energies, as illustrated by  Fig.~\ref{fig5}.
Thus  we  can   also  put  $\psi_{\varepsilon}   (\lambda)  \simeq
\psi_{\varepsilon} (\lambda_0)$ on the r.h.s. of Eq.~(\ref{4.5}).
As a result, both $c_{\lambda}$ and $\psi_{\varepsilon}  (\lambda)$
can be extracted  out the integral.   Using Eq.~(\ref{3.1.1})  we
immediately get $\delta T(t)  \propto e^{-\lambda_0
(t-t_0)}/(t-t_0)$, which coincides with Eq.~(\ref{3.2.8})
at $t - t_0 \gg \lambda_{1}^{-1}$.  Note that the above derivation
is based on  the particular spectrum  ($\lambda =0$ $+$  continuum
$\lambda_0 \leq  \lambda <  \infty$) of  the linearized  collision
integral and  has no  analogy in  the relaxation  kinetics due  to
particle-particle interaction.   In  the latter  case the fivefold
degenerate  eigenvalue  $\lambda  =  0$  is  separated  from   the
continuous  spectrum  by  a  set  of discrete isolated eigenvalues.
\cite{Cercignani88}

\section{DISCUSSION}
\label{sec5}

    In     order     to     test     the     generic      solution
(\ref{3.2.2})-(\ref{3.2.9})   we    model   the    phonon-assisted
relaxation of excitons within the initial kinetic Eq.~(\ref{1.1})
reduced to  energy space [Eqs.~(\ref{2.1})-(\ref{2.2})].   An
adaptive  inhomogeneous   grid  with   100  -   200  points   for
$\varepsilon$  is  used  to   cover  the  close  vicinity   of  the
ground-state  mode  $\varepsilon=0$  (the  maximum  value  of  the
dimensionless energy is $\varepsilon_{\rm  max} = 20$).   Equation
(\ref{2.2}) is evaluated by a fourth-order Runge-Kutta integration
routine  with  the  time  step  $0.001-0.01  \tau_{\rm  sc}$.   In order to
calculate integrals on the  r.h.s. of Eq.~(\ref{2.2}) we  perform
spline-approximation for  $f_{\varepsilon}(t)$ at  every iterative
step.

\begin{figure}
\includegraphics{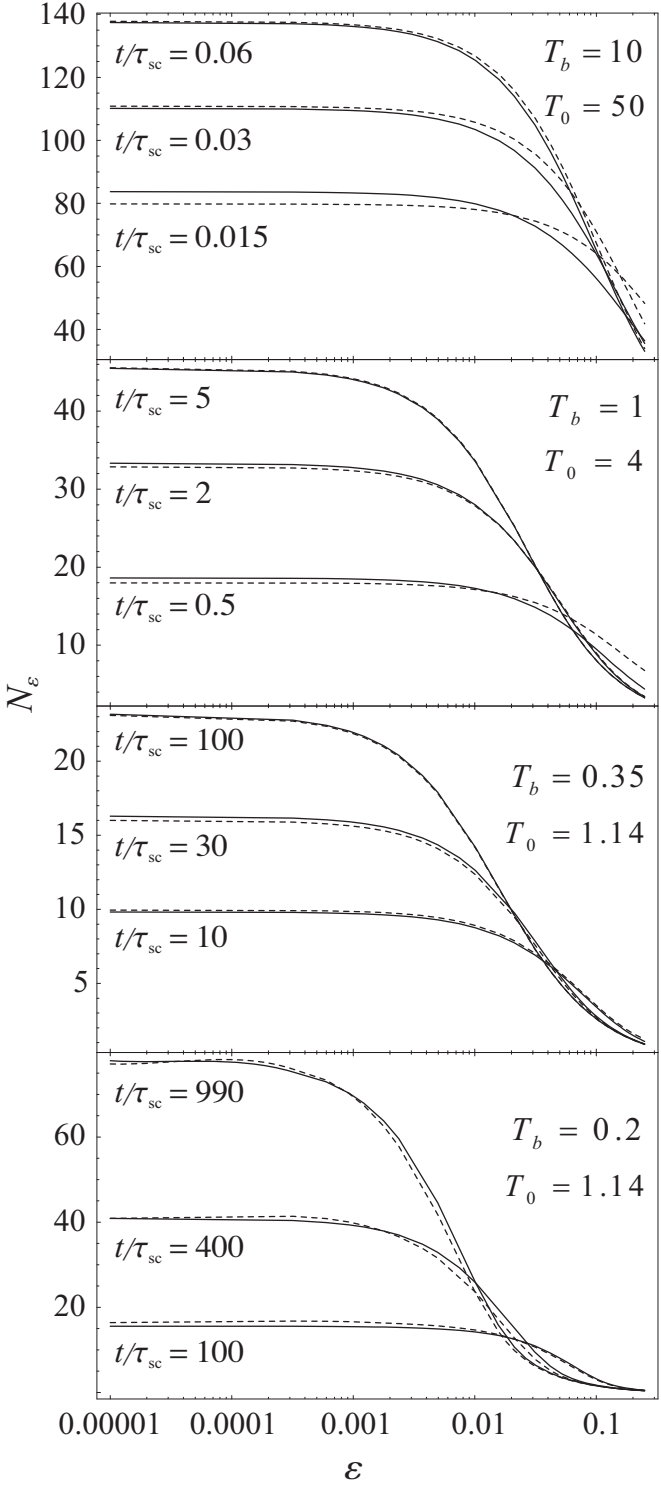}
\caption[Figure 6]{
    Evolution of  the distribution  function $N_{\varepsilon}$  at
the  adiabatic  stage  calculated  for  various  sets of the control
parameters, $T_b$ and $T_0$.  Solid lines correspond to  numerical
evaluation of  Eq.~(\ref{2.2}); dashed  lines are  obtained using
the generic solution given by Eq.~(\ref{3.2.5}).
}
\label{fig6}
\end{figure}

    In numerical simulations  we use the  dimensionless temperatures
$T_b$ and $T_0$  and measure time  in $\tau_{\rm sc}$.
This makes our  results  suitable  for  various  QWs 
and sets of the control parameters, provided that 
$E_0$, and $\tau_{\rm sc}$ are specified.  
In Fig.~\ref{fig6}    time    evolution    of    the    distribution
$N_{\varepsilon}(t)           =N^{0}_{\varepsilon}+            T_b
(N^{0}_{\varepsilon})' f_{\varepsilon}(t)$ as a numerical solution
of  Eq.~(\ref{2.2})  is  compared  with the corresponding generic
solution (\ref{3.2.2})-(\ref{3.2.9}) relevant to $T_b  < T_0$.  All  plots
demonstrate  an   excellent  agreement   between  analytical   and
numerical solutions.   Note,  that at  high $T_b$,  e.g., $T_b=10$
(see the top plot in Fig.~\ref{fig6}), the thermalization time $\tau_{\rm
th}  =  \lambda_{0}^{-1}$  achieves  the  limit $\tau_{\rm th} \ll
\tau_{\rm sc}$, and the relative duration of the
adiabatic stage estimated in terms of $t_c$ (duration of the first transient)
becomes smaller than that at $T_b \lesssim 1$.
In  this  case  the influence of the initial
distribution  $N_{\varepsilon}(t=0)$ slightly  affects  the  calculated
evolution at the
beginning of the adiabatic stage, and the analytical solution fits
numerically simulated data a little  worse as can be seen  for the
distribution functions at  $t= 0.015\tau_{\rm  sc}$.   In contrast,  at low
$T_b$ the relaxation kinetics at the adiabatic stage is slow.  For
example, at $T_b \leq 0.25$  it lasts more than $1000$  scattering
times, where typical  values of $\tau_{\rm  sc}$ in 
GaAs/AlGaAs coupled  QWs are on a scale of tens nanoseconds.

%

\begin{figure}
\includegraphics{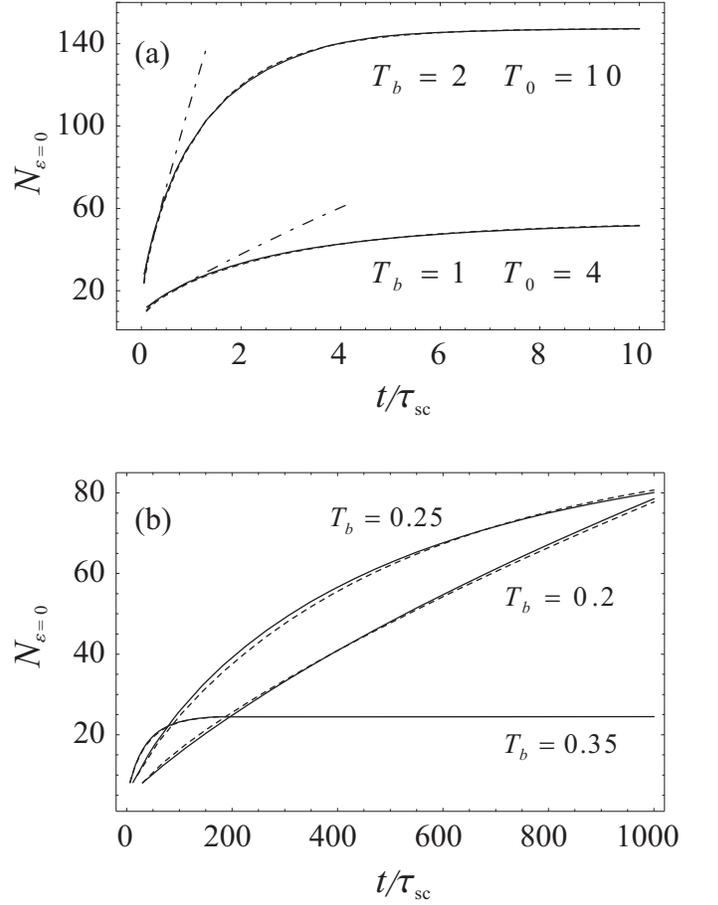}
\caption[Figure 7]{
    Population dynamics of the ground  state mode
$N_{\varepsilon=0}=N_{\varepsilon=0}(t)$  
calculated  for different  bath temperatures  at
$T_0=4,10$ (a), and  $T_0=1.14$ (b).  Similarly to Fig.~\ref{fig6},
the solid and dashed curves are calculated by using
Eqs.~(\ref{2.2}) and (\ref{3.2.5}), respectively.  Dash-dotted
lines  correspond   to  time   dependences  $N_{\varepsilon=0}(t)$
calculated using approximate Eq.~(\ref{5.1}).
}
\label{fig7}
\end{figure}

    In Fig.~\ref{fig7} we also  compare time dependences for  the
ground-state                    mode                    population
$N_{\varepsilon=0}=N_{\varepsilon=0}(t)$   obtained    numerically
(solid lines) and analytically (dashed lines).  Again, this figure
shows    that    at    $\varepsilon=0$    the   generic   solution
(\ref{3.2.2})-(\ref{3.2.9}) reproduces the adiabatic stage of  the
phonon-assisted relaxation kinetics very well.   Within
the time interval $0 \le t-t_0 \lesssim \tau_{\rm th}$ the generic
solution  yields  the  following  simple approximation  for the adiabatic
kinetics into the ground state mode:
\begin{equation}
N_{\varepsilon=0}(t) =N_{0}^{\rm in} [1 + \chi (t-t_0)]^{\nu}.
\label{5.1}
\end{equation}
    Parameters  $\chi$  and  $\nu$  can  be found comparing series
expansions of Eq.~(\ref{5.1})  and (\ref{3.2.5}) about  the point
$t=t_0$.  In such  a way we get  $\chi=(c_{1}^{2}-c_{2} N_{0}^{\rm
in})/(c_{1} N_{0}^{\rm  in})$ and  $\nu=c_{1}^{2}/(c_{1}^{2}-c_{2}
N_{0}^{\rm in})$, where time-independent constants $c_1$ and $c_2$
are given by
\begin{eqnarray}
&c_{1}=& (N^{0}_{\varepsilon=0} - N_{0}^{\rm in})
[\lambda_0 + \delta T_0 \xi_{1}(0)]
+T_b  \delta T_0 \eta(0) (N^{0}_{\varepsilon=0})' ,
\nonumber\\
&c_{2}=&\frac{1}{2}
\biglb (
(N_{0}^{\rm in} - N^{0}_{\varepsilon=0})
\nonumber\\
&& \times
\{
2 \lambda_{0}^{2} + \delta T_0 \xi_{1}(0)
[5 \lambda_{0}+\lambda_1+2  \delta T_0 \xi_{1}(0)]
\}
\nonumber\\
&& - T_b \delta T_0 \eta(0) (N^{0}_{\varepsilon=0})'
[
3 \lambda_0 +\lambda_1 + 2  \delta T_0 \xi_{1}(0)
]
\bigrb) .
\label{5.2}
\end{eqnarray}
    Time   dependences   of   the   ground-state  mode  population
$N_{\varepsilon=0}(t)$ calculated at $T_b =1$ and $T_b = 2$  using
Eq.~(\ref{5.1}) are shown in Fig.~\ref{fig7}(a) with dash-dotted
lines.  At $t - t_0 > \tau_{\rm th}$ the approximation (\ref{5.1})
violates  because  the phonon-assisted relaxation kinetics
starts to become exponential.

%

    As we have shown in Sec.~\ref{sec4}, for a given concentration
of QW excitons $\rho_{\rm 2D} \lesssim 10^9$ cm$^{-2}$ the
thermalization kinetics slows down  with decreasing  
$T_b/T_0$, i.e., with development of quantum statistics.  
By  increasing  both  temperatures,  $T_0$ and $T_b$, and
keeping unchanged the ratio
$T_0/T_b  \gg  1$  one  can simultaneously avoid the above
bottleneck effect in relaxation and achieve high population of the
ground-state mode, $N_{\varepsilon=0}\gg 1$.  However, in this case  the
concentration of excitons $\rho_{\rm 2D} \propto T_0$ increases as
well, so that  exciton-exciton interaction eventually  becomes the
main mechanism of relaxation in GaAs/AlGaAs coupled or ZnSe single QWs.

\begin{figure}
\includegraphics{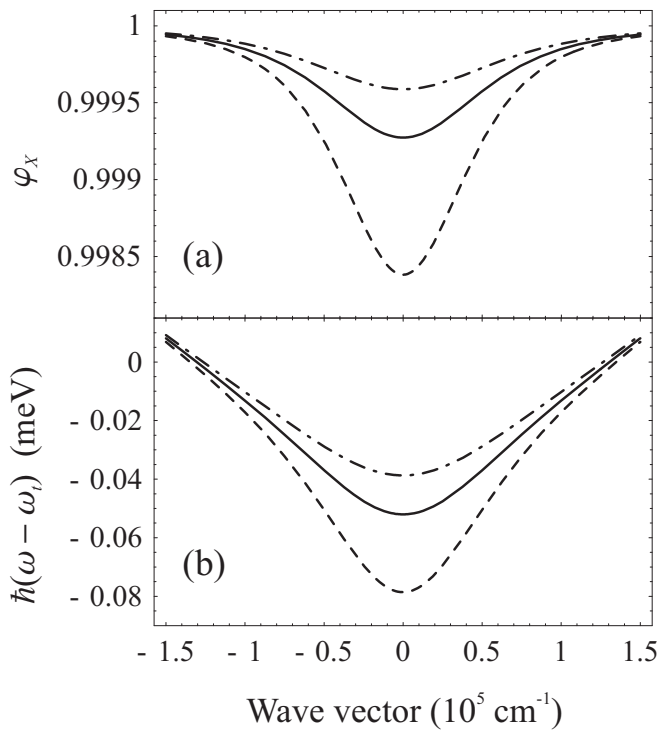}
\caption[Figure 8]{
    Possible  design  of  GaAs-based  microcavities:   (a) excitonic
component $\varphi_{X}=\varphi_{X}({\textbf k_{\|}})$  
of the  MC polariton eigenstate;  and (b) the lower  branch polariton
dispersion $\hbar(\omega-\omega_t)$.  Detuning $\delta=50$  meV 
(dashed lines), $75$ meV (solid lines), and  $100$
meV  (dash-dotted  lines).  The energy of ground-state QW  excitons
is given by $E_{{\textbf k_{\|}}=0} = \hbar \omega_t =1.522$ eV.
}
\label{fig8}
\end{figure}

In contrast, in high-quality GaAs-based microcavities with a relatively large
positive detuning  $\delta =  \hbar(\omega_0 -  \omega_t)$ between the 
cavity   mode   ($\hbar\omega_0$)   and   QW   exciton  line
($\hbar\omega_t$) the LA-phonon-assisted kinetics remains dominant even at
relatively high degeneracy temperatures $T_0 \gg T_b > 1$. In these systems,
by changing the detuning $\delta$ within the band $\delta = 50 - 100$ meV
one can design an effective in-plane mass $M_x$ so that the bottleneck
effect in phonon-assisted scattering, due to the low density of states
$\propto M_x$, is already relaxed, whereas the  degeneracy  temperature 
$T_0  \propto \rho_{\rm 2D}/(E_0  M_x)$ is still relatively high. Indeed, 
in  such  MCs the lower polariton branch gives rise to the in-plane
translational mass much smaller than  the mass of optically-undressed QW 
excitons.  In the meantime the excitonic component $\varphi_{X}$ of the
microcavity polaritons is already very  high, $\varphi_{X} \gtrsim 
0.999$, resulting in the long optical decay (in the $z$-direction)
lifetimes on a few nanosecond time scale. This  is illustrated in
Fig.~\ref{fig8}(a), where the detuning $\delta$ is equal to $50$, $75$, and
$100$ meV. The    corresponding  polariton (exciton) masses are given by
$M_{x} = 0.023 m_0$, $0.050 m_0$, and $0.081 m_0$, respectively. The
relevant lower-branch polariton dispersions are plotted in 
Fig.~\ref{fig8}(b). Because the  energy  $E_0$  is only on a $0.01$ meV
energy scale, the parabolic approximation of the lower-branch dispersion
curves is valid for low-temperature relaxation kinetics. Thus the
LA-phonon-assisted thermalization of the exciton-like MC polaritons can
indeed be modelled by the kinetic Eqs.~(\ref{2.1})-(\ref{2.2}). Time  evolutions  of  the  
exciton distribution $N_{\varepsilon}(t)$, which are typical for the
proposed design of GaAs-based MCs, are shown in Fig.~\ref{fig6} (see the
plots with  $T_b  =  1$,  and  $10$). Figure \ref{fig7}(a) illustrates
the corresponding population dynamics of the ground-state mode,
$N_{\varepsilon=0}=N_{\varepsilon=0}(t)$.

%

\section{CONCLUSIONS}
\label{sec6}

    In  this  paper  we  have  studied  thermalization kinetics of
statistically  degenerate  QW  excitons  coupled  to  thermal bulk
acoustic phonons.   For  concentrations of  QW excitons $\rho_{\rm
2D} \lesssim 10^9$ cm$^{-2}$ the particle-particle interaction  in
GaAs or ZnSe QWs can be neglected in comparison with QW exciton --
bulk acoustic phonon  scattering, and the  thermalization kinetics
from  an  initial  distribution  of  QW  excitons  at $t=0$ occurs
through the nonequilibrium   states.     The  following   conclusions
summarize our results.

    (i) For the  case of well-developed  Bose-Einstein statistics,
when $T_b < T_0$ so  that $N_{\varepsilon=0} > 1$, the  relaxation
kinetics of QW excitons  coupled to thermal bulk  acoustic phonons
is given by the following  scheme.  Within a few  characteristic
scattering  times  the  correlation  of  the distribution function
$N_{\varepsilon}(t)$  with   the  initial   $N_{\varepsilon}(t=0)$
disappears, and  the subsequent thermalization of  QW excitons  is
described in  terms of  the adiabatic stage  of relaxation.
The adiabatic stage is characterized by the start time  $t_c$,
which absorbs a gross  information about the initial  distribution
$N_{\varepsilon}(t=0)$, and  by the  parameter $\lambda_0$,  which
depends only upon the bath and degeneracy temperatures, $T_b$  and
$T_0$.   At the  beginning of  the adiabatic  stage, i.e., for the
time  domain  $0  \leq t - t_c  \lesssim  \lambda_0^{-1}$, one has
$|N_{\varepsilon \leq 1/4} - N_{\varepsilon  \leq  1/4}^0| \simeq
N_{\varepsilon   \leq   1/4}^0$,  and  the
thermalization kinetics is strongly nonexponential, with $\delta T
\propto 1/t$  and $N_{\varepsilon=0}  \propto (1+  \chi t)^{\nu}$.
At large times,  when the deviation  of the system  from the final
equilibrium  state  is   already  small  ($|N_{\varepsilon}(t)   -
N_{\varepsilon}^0|/N_{\varepsilon}^0 \ll 1$), the adiabatic  stage
of the phonon-assisted thermalization becomes exponential, $\delta
T  \propto  e^{-\lambda_0  t}$,  and  can  be described within the
linearized kinetic equation.

    (ii) The linearized LA-phonon-assisted kinetics of QW excitons
is formulated in terms of the Fredholm integral  Eq.~(\ref{4.1}).
The  eigenvalues  $\{  \lambda  \}$  of the collision integral are
given by the continuous spectrum $\lambda_0 \leq \lambda < \infty$
and  the  isolated  eigenvalue  $\lambda  =  0$.    The   marginal
eigenvalue   $\lambda_0   =   \lambda_0(T_b,T_0)$  determines  the
thermalization time at $t \rightarrow \infty$ by $\tau_{\rm th}  =
\lambda_0^{-1}$.   In dependence  on the  two control  parameters,
$T_b$ and  $T_0 \propto  \rho_{\rm 2D}$,  the thermalization  time
achieves  two  limits:    $\tau_{\rm  th}  \ll  \tau_{\rm sc}$ and
$\tau_{\rm   th}   \gg   \tau_{\rm   sc}$.     The  eigenfunctions
$\psi_{\varepsilon}(\lambda)$    of    the    collision   integral
(\ref{4.1}) are  smooth integrable  functions with  three isolated
critical  points  in  energy  space.    The  critical  points   of
$\psi_{\varepsilon}(\lambda)$ give rise to  a first order pole,  a
logarithmic singularity, and a continuous step-like jump.

    (iii) Because the LA-phonon-assisted kinetics becomes dominant
only  at  small  concentrations  of  QW  excitons,  $\rho_{\rm 2D}
\lesssim  10^9$  cm$^{-2}$,  nonclassical  statistics  of quasi-2D
excitons  in  ZnSe  or  GaAs  QWs  develops  at  very  low   bath
temperatures  $T_b  <  1$  K.  The  proposed  design of GaAs-based
microcavities with  a relatively  large positive  detuning between
the cavity mode and  QW exciton line, $\hbar(\omega_0  - \omega_t)
\geq    50$    meV,    allows    us,    however,   to   build   up
$N_{\varepsilon\simeq0} \gg 1$ by means  of QW exciton -- bulk  LA
phonon scattering in much  more favorable conditions, i.e.,  $T_b
\gtrsim 1$ K and $\tau_{\rm th} \ll \tau_{\rm sc}$.

%
\begin{acknowledgments}
    We  appreciate  valuable  discussions  with  L.V.    Butov, V.
Fal'ko, and  M.S. Skolnick.   Support  of this  work by  the EPSRC
(U.K.) is gratefully acknowledged.
\end{acknowledgments}

\appendix*
\section{Temperature law}
\label{appendb}

    In Appendix we give the expressions for the parameters
and functions
used in Eqs.~(\ref{3.3.1}), (\ref{3.3.6}), and (\ref{3.3.7}).
The parameters $\alpha_0$, and $\beta$ arise from
those  terms in the collision integral which
contain $f_{\varepsilon \geq 1/4}(t)$ and $f^{2}_{\varepsilon \geq
1/4}(t)$, respectively.  Collecting such terms together one gets
\begin{eqnarray}
\alpha_0 &=&
\frac{4}{\tau_{\rm sc} \varepsilon_c}
\left[
\varepsilon_c \int^{\varepsilon_c}_{\varepsilon_c -1/4}
F_{\rm S}(\varepsilon_c, \varepsilon)
\left(
1+ n^{\rm ph}_{\varepsilon} + N^{0}_{\varepsilon_c - \varepsilon}
\right) \varepsilon d \varepsilon  \right.
\nonumber\\
&&+  \int^{\varepsilon_c -1/4}_{0}
F_{\rm S}(\varepsilon_c, \varepsilon)
\left(
1+ n^{\rm ph}_{\varepsilon} + N^{0}_{\varepsilon_c - \varepsilon}
\right) \varepsilon^{2} d \varepsilon
\nonumber\\
&& - \left.
 \int^{\infty}_{0}
F_{\rm AS}(\varepsilon_c, \varepsilon)
\left(
n^{\rm ph}_{\varepsilon} - N^{0}_{\varepsilon_c + \varepsilon}
\right) \varepsilon^{2} d \varepsilon
\right] ,
\label{b1}
\end{eqnarray}
and
\begin{eqnarray}
\beta &=&
\frac{4}{\tau_{\rm sc} T_b}
\left[
\int^{\varepsilon_c -1/4}_{0}
F_{\rm S}(\varepsilon_c, \varepsilon)
(N^{0}_{\varepsilon_c - \varepsilon})'
(\varepsilon_c - \varepsilon)  \varepsilon d \varepsilon  \right.
\nonumber\\
&& - \left.
 \int^{\infty}_{0}
F_{\rm AS}(\varepsilon_c, \varepsilon)
(N^{0}_{\varepsilon_c + \varepsilon})'
(\varepsilon_c + \varepsilon)  \varepsilon  d \varepsilon
\right] .
\label{b2}
\end{eqnarray}
    The function $\alpha_1(t)$ stems  from
the   terms   proportional   to   $f_{\varepsilon   \geq   1/4}(t)
f_{\varepsilon < 1/4}(t)$ and is given by
\begin{equation}
\alpha_1(t) =
\frac{4 T_b}{\tau_{\rm sc}} \int^{1/4}_{0}
F_{\rm S}(\varepsilon_c,\varepsilon_c-\varepsilon)
(N^{0}_{\varepsilon})'
f_{\varepsilon}(t)
(\varepsilon_c-\varepsilon) d \varepsilon.
\label{b3}
\end{equation}

   The parameter ${\tilde\alpha}$ from Eq.~(\ref{3.3.3})
is obtained by collecting  all the terms $\propto \delta T(t)$: 
\begin{eqnarray}
{\tilde\alpha} &=&
- \int^{\varepsilon_c}_{\varepsilon_c -1/4}
F_{\rm S}(\varepsilon_c, \varepsilon)
 \left(
1+ n^{\rm ph}_{\varepsilon} + N^{0}_{\varepsilon_c - \varepsilon}
\right)
\frac{\eta(\varepsilon_c - \varepsilon)}
{\xi_0(\varepsilon_c - \varepsilon)}
\varepsilon d \varepsilon
\nonumber\\
&&- \left(\alpha_0 \tau_{\rm sc} \varepsilon_c \right) /
\left(4 T^{2}_{b}
\right).
\label{b4}
\end{eqnarray}

  When  deriving  Eq.~(\ref{3.3.7})  from Eq.~(\ref{3.3.6}) we
first  put   $\varepsilon  =   0$  everywhere   in  the  integrand
(\ref{3.3.6})  except   the  exponent.    By  changing  the
integration variable $\varepsilon$ to $\xi= \xi_{0}(\varepsilon)$
and putting $\varepsilon  = 0$ we derive the  asymptotic Eq.~(\ref{3.3.7}), 
valid  in the limit   $t   \to   \infty$.  In this equation  ${\tilde\gamma}_0$  is a
time-independent  pre-factor given by
\begin{equation}
{\tilde\gamma}_0 =
F_{\rm S}(\varepsilon_c, \varepsilon_c)
 \left(
1+ n^{\rm ph}_{\varepsilon_c} + N^{0}_{\varepsilon=0}
\right)
\frac{\varepsilon_c  f_{\varepsilon=0}(t_0) }{ \xi_{0}'(0)}.
\label{b5}
\end{equation}
%



\end{document}